\documentclass[journal,transmag]{IEEEtran} 
\ifCLASSINFOpdf
\usepackage{graphicx}
\else
  \usepackage[dvips]{graphicx}
\fi
\begin{document}
%
\title{ Transparent Negative Index of Refraction Metamaterial Using a Wire Array in a Magnetic Host}


\author{\IEEEauthorblockN{Mohamed Zein Radwan\IEEEauthorrefmark{1},
Graeme Dewar\IEEEauthorrefmark{1}}
\IEEEauthorblockA{\IEEEauthorrefmark{1}Department of Physics and Astrophysics, University of North Dakota, Grand Forks, ND 58202 USA}
\thanks{ 
Corresponding author: G. Dewar (email: graeme.dewar@ndus.edu).}}

%


\IEEEtitleabstractindextext{%
\begin{abstract}
We have made  measurements of microwave transmission 
over the 12 - 18~GHz range and 
through a simple metamaterial exhibiting a negative index of refraction. 
The metamaterial consisted of an array of wires cladded in dielectric  embedded in a magnetic ferrite. The ferrite replaced the cut-ring structure usually used to create the negative permeability.
The dielectric cladding decoupled the ferrite from the wires, thereby allowing the wire array permittivity to be simultaneously negative with the permeability.  The simplicity of the design allows for miniaturization of potential microwave devices based on this metamaterial. 
\end{abstract}

}

\maketitle

\IEEEdisplaynontitleabstractindextext

%
\IEEEpeerreviewmaketitle

\section{INTRODUCITON} 
%
%
%
%
 
Since the first demonstration of a two-dimensional metamaterial with a negative index of refraction [Smith 2000], there have been many advances in tailoring the electromagnetic response of man-made structures, including extending demonstration of the phenomenon from the microwave frequencies to the optical range [Shalaev 2007].   
Besides verifying the fact that a light ray passing through a prism having a negative index of refraction bends in the opposite direction to that found with an ordinary prism [Shelby 2001],  sub-wavelength imaging with a spherical aberration free lens [Pendry 2000]
is well understood.  In addition,  modification of the permittivity $\epsilon$ 
and permeability 
$\mu$ 
in the region of space surrounding an object has 
resulted in phenomena ranging from   
 ``cloaking'' 
[Pendry 2006, Schurig 2006, Cai 2007, Wood 2009]
of that  object to having the object become a super scatterer 
[Yang 2008, Wee 2009].
Indeed, the concepts of cloaking and super scattering  have been extended beyond Maxwell's equations to the diffusion equation and the possibility of creating metamaterials for mass separation 
[Guenneau 2013, Restrepo-Fl\'orez 2016].
Other predictions [Veselago 1968] regarding  reverse Cherenkov radiation 
[Chen 2011a, Duan 2017]
and the reverse Doppler effect [Chen 2011b] have been experimentally verified.  However, the
reverse Casimir effect [Veselago 1968, Leonhardt 2007] remains speculative.

The key ingredient to guarantee a negative index of refraction whereby the light wave's phase fronts move in the direction opposite  to which energy propagates 
is that both the  permeability and  permittivity are negative.  Implicit in this is that the losses in the medium, usually described by 
the imaginary parts of the permeability and permittivity,  are small;  this ensures that the medium is reasonably transparent.  The negative response functions are typically obtained with resonant structures driven at frequencies somewhat above their natural ones. Then the responses are more than 90$^o$ out of phase with the drives and, provided that responses are large enough, the permittivity and permeability are negative.

Only negative permeability is required to produce a material which has a negative phase velocity.  
See, for example, (21) and (25) in  
Cochran et al.~[1977].
If the permittivity is positive with $\mu < 0$ then the index of refraction calculated from Maxwell's equations is an imaginary number, hence the medium is highly reflective and  does not permit light to freely propagate through it. Atypically, an array of ferrite rods [Gu 2013] alone can have a negative effective permittivity simultaneously with $\mu < 0$ but this is not the usual case.
Also,
a metallic ferromagnetic conductor can have 
$\mu < 0$ and hence $n < 0$ [Pimenov 2007] but the inevitable attenuation associated with ohmic losses destroys the medium's transparency. 
In the simplest case a
negative permittivity coupled with a negative permeability, both accompanied by low losses,  
results in a negative real valued  index of refraction and the 
negative index material (NIM)
is transparent.

The first demonstrated NIM used an array of conducting wires 
for which permittivity was negative for frequencies less than the array's plasma frequency.  The negative permeability was supplied by an array of cut-rings with a resonant frequency slightly below the frequencies of interest.  In the experiments described here the NIM was fabricated with a nonconducting
magnetic host 
which supplied the negative permeability and a wire array, suitably decoupled from the magnetic host, supplied the negative permittivity.
This is a much simpler structure to fabricate than the wire/cut-ring arrays.   This metamaterial has the added features that the range of microwave frequencies over which it exhibits a negative index of refraction can be tuned by changing the biasing magnetic field acting on it and the negative index of refraction is relatively isotropic for microwaves propagating in the plane perpendicular to the ferrite's magnetization direction.

\section{METAMATERIAL DESIGN AND FABRICATION} 

Simply embedding an array of wires in a non-conducting magnetic material does not lead to a NIM [Pokrovsky 2002, Dewar 2002] because material having $\mu < 0$ in close contact with the wires causes the wire array to acquire a positive permittivity. The cure
to restoring $\epsilon <0$ is to surround
the wires with a non-magnetic dielectric [Dewar 2002, Dewar 2005a]. 
For these structures the permittivity is determined from the (angular) plasma frequency $\omega_p$ by 
$$
\epsilon = \epsilon_0\left({1 - \frac{{\omega_p^2}}{ \omega^2}}\right)
\eqno{(1)}
$$ 
and is negative for frequencies $\omega < \omega_p$.
The usual expression for the plasma frequency of a metal,
$$
\omega_p^2 = \frac{ne^2}{m_{\rm eff}\epsilon_0},
\eqno{(2)}
$$
is modified in two ways for the wire array [Pendry 1996].  First, the electron number density $n$ for the array of conducting wires is diluted by the volume between the wires and is much smaller than for a conductor filling all space.  Second, the electron effective mass $m_{\rm eff}$ is much enhanced by the wires.  The effective mass relates the work done by the electric field on electrons to change their kinetic energy $m_{\rm eff} v^2 /2$;   for electrons in the ionosphere the effective mass is just the usual electron mass while for electrons in a metal the effective mass is related to the metal's band structure.  For our wire array, an electric field that increases the speed $v$ of the electrons 
(and increases the electric current) 
creates an augmented local magnetic field around each wire. 
The energy associated with this near-zone magnetic field is much greater than the electronic kinetic energy, thus
$m_{\rm eff} \propto L$, where $L$ is the inductance of the wire.  But if $\mu < 0$ then $L < 0$, leading to a negative $m_{\rm eff}$ in (2) and a positive $\epsilon$ in (1), thus destroying the $n<0$ property.  Surrounding each wire with a non-magnetic dielectric creates a volume in which the inductive energy is positive and, provided the magnitude of the (negative) permeability in the magnetic host is not too large, the permittivity is negative, thus restoring the $n<0$ property.  Calculations show [Dewar 2002] that choosing the outer radius of the cladding to be the geometric mean of the wire radius and the array lattice constant provides an suitable frequency range over which both $\mu$ and $\epsilon$ are negative.

\begin{figure}[!t]
\includegraphics[scale=0.240]{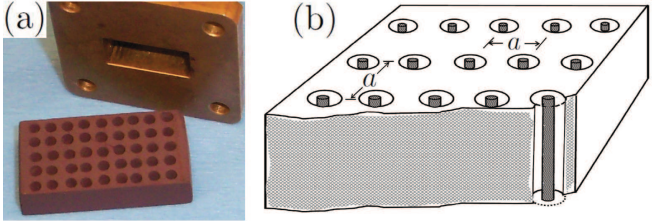}
\caption{\label{fig:sketch}
(a) Photograph of the ferrite block with a square array of holes drilled thorough it.  For the transmission experiments the holes were were threaded with copper wires cladded in Teflon$^{\rm TM}$ tubing and the block was inserted into the section of
waveguide shown.
(b) 
Cut-away sketch of the nickel zinc ferrite block.  The cladded wires formed a $9 \times 5$ square array with lattice constant $a = 3.0$~mm.
}\end{figure}

The structure described here offers advantages over other successfully demonstrated  $n < 0$ metamaterials employing a ferrite for the $\mu < 0$ property.  The multi-layer wire and ferrite rod or plate configurations  
[He 2009, Zhao 2007, Zhao 2009, Bi 2013, Bi 2014]
is somewhat more complicated to construct and can result in a highly anisotropic index of refraction [Rachford 2007].

\begin{figure}[!t]
\includegraphics[scale=0.3650]{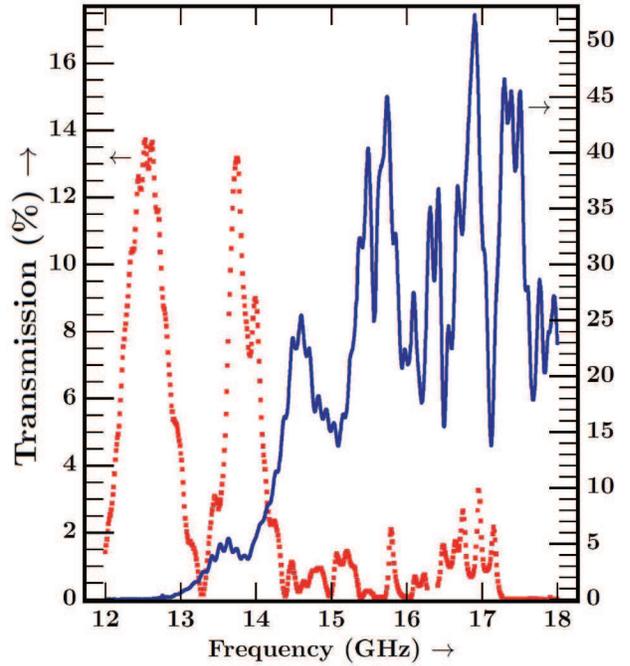}
\caption{\label{fig:maindata} Transmitted power in percent versus frequency of incident microwaves for ferrite blocks in waveguide.  The red square symbols represent the transmission data acquired for the ferrite block/wire array in an applied field of 
$2.4\times 10^4$~A/m
(300~Oe).  The blue filled line represents the transmission through a similarly sized ferrite block
which had no wires or holes in it and which was in an applied field of 
$8.0 \times 10^4$~A/m (1.0~kOe). The scale for the  filled line has been compressed by a factor of 3.0 for ease of comparison.
Due to the negative demagnetization factors associated with the holes, the internal magnetic fields in both blocks was approximately the same.  For both samples, the permeability was negative at frequencies less than 16~GHz.  
The significant transmission  for  the ferrite block/wire array in the 12~-~14~GHz range demonstrates transparency for $n < 0$.}
\end{figure}

Fig.~\ref{fig:sketch}  is a photograph of the ferrite block/cladded wire used in the microwave transmission measurements described here together with a schematic of the array of wires threading the block.  
Ferrite blocks were fabricated by cold pressing ($4.0 \times 10^4$~N) powdered starting material (19\% by weight NiO, 66\% Fe$_2$O$_3$, 14\% ZnO, and 1\% MnO$_2$) in cylindrical molds 39~mm in diameter.  Rectangular blocks were cut from from the pressed material and sintered at $1000^o$C for five hours with heating and cooling rates of 100~C$^o$/hour.  
This low sintering temperature resulted
in blocks which were structurally strong but could be drilled with ordinary drill bits and
were easily shaped by sawing and sanding into rectangular parallelepipeds of final dimensions of
$30 \times 15.7 \times 7.9$~mm$^3$. The block shown in Fig.~\ref{fig:sketch}
had
a $5 \times 9$~square array of 1.7~mm diameter holes, lattice constant 3.0~mm, drilled into it.  
Each hole was threaded with a 0.29~mm diameter copper wire cladded with Teflon$^{\rm TM}$ tubing having an inner diameter of 1.0~mm and outer diameter 1.6~mm.

\section{TRANSMISSION MEASUREMENTS} 

Fig.~\ref{fig:maindata} displays the measured microwave transmission through the ferrite block 
with a wire array threaded through it together with the transmission through a similarly sized ferrite block which did not have an array of holes or wires in it.  
The ferrite blocks each 
filled the lateral dimensions of  a section of WR-62 wave guide 
and were mounted in the gap of a 9-inch Varian electromagnet.  Additional blocks of ferrite which had undergone the same preparation history as the samples were mounted in the magnet's gap.  These blocks constituted part of a magnetic circuit that, in the absence of any holes in the sample, limited the static demagnetization factor acting on the samples to less than $10^{-2}$.
The essential difference between the two measurements is that the negative permittivity of the wire array allowed significant transmission at frequencies below $14$~GHz for which the permeability  was negative (and $n < 0$) and reduced the transmission at frequencies $> 16$~GHz for which $\mu > 0$.  
The sample without wires behaved in the opposite manner, having significant transmission at higher frequencies with both response functions positive ($n>0$).  This is a clear demonstration that our sample exhibited a negative index of refraction in a manner similar to the demonstration [Smith 2000].

The maximum transmission of our $n<0$ metamaterial is approximately 1/3 that of the ferrite.  Since both samples had measured reflectivity's on the order of 50\%, we attribute the lower transmission to resistive losses in the wire array.  The internal magnetic fields for the two measurements shown in Fig.~\ref{fig:maindata} are comparable because the holes in the ferrite gave rise to a significant negative demagnetization factor.  We calculated the demagnetization factor to range from -0.16 near the center of our sample to -0.05 near a corner.  This inhomogeneous demagnetization factor may have also led to a decrease in the observed transmission relative to the ferrite block with no wires or holes in it.

\section{PROPAGATION CONSTANT CALCULATION} 

\begin{figure}[!t]
\includegraphics[scale=0.3250]{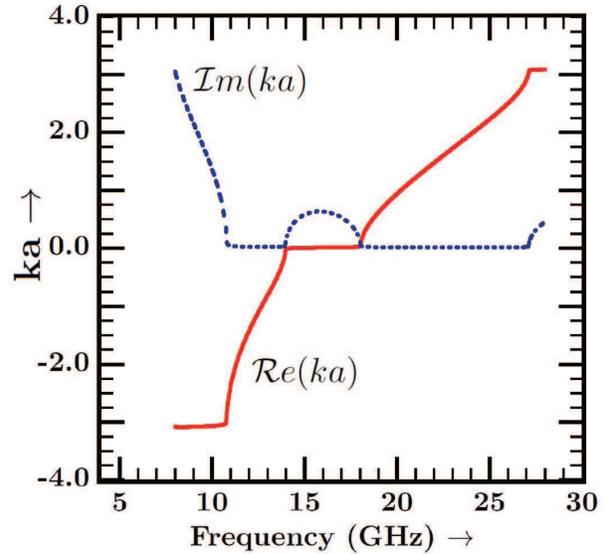}
\caption{\label{fig:calc}Plot of the propagation constant $k$  versus frequency according to the calculation outlined in Dewar [2005b]). The ferrite/wire array is assumed infinite in extent and $k$ has been scaled by the lattice constant $a$ of the wire array.  Propagation is along the $(1, 0)$ direction and $k$ has been adjusted 
to account for the cut-off of long wavelength radiation in waveguide;
it is the component of $k$ parallel the waveguide transmission direction for the TE$_{1,\,0}$
mode.
A negative index of refraction corresponds to the real
part of $k$ negative while the imaginary part, representing dissipation, is small. Between 11 and 14~GHz the index of refraction ranges from $-4.5$
to approximately $0$.
Immediately below 11~GHz there is a Bragg reflection.
The band gap between 14 and 18~GHz is due to two causes: 1)~$k$ is below the cut-off for propagation in the waveguide and 2)~the permittivity of the wire array is negative while the  permeability averaged over a cell of the array, including the hole and wire, 
is small and positive.  Between 18 and 27~GHz the infinite ferrite/wire array is calculated to exhibit a positive index or refraction but has a Bragg reflection above 27~GHz.
 }
\end{figure}

Fig.~\ref{fig:calc} is a plot of the calculated [Dewar 2005b] propagation constant, $k$, versus frequency for waves of the form $\exp(i{\bf k\cdot r} - i\omega t)$ in an infinite medium. Note that $k$ plotted in Fig.~\ref{fig:calc} has been scaled by the lattice constant of the wire array.  The values of the magnetization ($M = (3.80\pm0.21) \times 10^5$~A/m)
and $g$-factor ($g = 2.006 \pm 0.049$) required in the calculation were obtained from ferromagnetic resonance measurements on a small cylindrically shaped sample of the ferrite.  The ferrite's intrinsic permittivity 
($\epsilon = (4.45 \pm 0.5)\epsilon_0$)
was determined from the spacing in frequency at $H = 8.8\times 10^5$~A/m ($\mu > 0$, $\epsilon > 0$) of  transmission modes [Barzilai 1958] for the ferrite block having no wires in it.

The relevant feature of Fig.~\ref{fig:calc} is that the real part of $k$ is negative for 11~GHz $\le f \le 14$~GHz and the imaginary part of $k$ (representing dissipation) is small.  
Over this frequency interval the index of refraction is calculated to fall in the range $-4.5 \le n \le 0$.
In the calculation $\mu_{\rm ferrite} < 0$ for $f < 16$~GHz.  
The relatively large transmission through our NIM sample falls in this 
frequency range.  
The ``band gap'' (large imaginary part of $k$ with the real part near 0) 
between 14 and 18~GHz 
is due to two causes. First, the magnitude of the ferrite's negative permeability is so small that the permeability averaged over a cell of the photonic lattice is positive and second, the $k$'s magnitude was reduced by a component equivalent to half a wavelength across the width of the waveguide representing the transmission cut-off for the waveguide.
The calculation also shows that, for $f < 11$~GHz, there is a Bragg reflection of the microwaves and transmission is curtailed.  In addition, the  negative permeability below 11~GHz is so large that it destroys the negative permittivity of the wire array, further enhancing attenuation of the microwaves.  Any transmission through our ferrite/wire array sample between 11 and 12~GHz could not propagate in the waveguide we used.  The transmission we observed through our ferrite with the wire array, and  shown in Fig.~\ref{fig:maindata}, is large for $k < 0$, equivalent to $n<0$, and small otherwise.

\section{CONCLUSIONS}  

With proper impedance matching a microwave device based on our NIM would have a 5~dB insertion loss.  Further improvement is possible by optimizing the wire array lattice constant and radii of the wires and holes.  A potential application could be a directional coupler capable of reversing the direction from which microwaves are coupled.  
Our sample exhibited a positive $n$ and transmission comparable to that shown in Fig.~\ref{fig:maindata} in an applied field of $6.0 \times 10^5$~A/m (7.5~kOe). 
All that need be done to reverse the input coupling direction 
with a similar material in a directional coupler
is change the biasing magnetic field on the ferrite.   More generally, our transparent NIM has a simpler structure than earlier NIM's based in rods (or wires) and cut-ring structures. 
The waveguide used to house our NIM could, for example, be easily  reduced to transverse dimensions of 5.0~mm by 1.0~mm.
Further, the operating frequency of our NIM can be adjusted by changing the biasing magnetic field acting on it, thus offering the prospect of more readily realized novel uses.

The success of using a magnetic material as one of the two crucial components of a NIM suggests that it is possible to simply achieve a NIM
at much higher frequencies.  
Ferrimagnets and antiferromagnets have a resonance in the terahertz or far infrared regime whereby  two magnetic sublattices exert torques on each other [Moorish 1965].  A semiconducting ferrite or antiferromagnet could have its plasma frequency tuned, perhaps by changing temperature, to the same terahertz frequency, thereby yielded simultaneously negative $\mu$ and $\epsilon$.  Such a material would have a narrow frequency range, perhaps only a few gigahertz wide, over which it was a NIM but this frequency interval could be tuned by an applied magnetic field.

\bigskip
\centerline{REFERENCES}
\medskip


\hfill\break
$\phantom{\,}$
\hskip-4.50truemm
Barzilai G,  Gerosa G (1958), 
``Modes in rectangular guides filled with magnetized ferrite,''
{\it Il Nuovo Cimento,} vol.~7, pp.~685--697,
doi: 10.1007/BF02781572.

\hfill\break
$\phantom{\,}$
\hskip-4.50truemm
Bi K, Zhou J, Zhao H, Liu X,  Lan C (2013), 
``Tunable dual-band negative refractive index in ferrite-based metamaterials,''
{\it Opt. Express,} vol.~21, pp.~10746--10752,
doi: 10.1364/OE21.010746.
\hfill\break

\noindent 
$\phantom{\,}$
\hskip-4.50truemm
Bi K, Guo Y, Zhou J, Dong G, Zhao H, Zhao Q, Xiao Z, Liu X,  Lan C (2014), 
``Negative near zero refraction metamaterials based on permanent magnetic ferrites,''
{\it Sci. Rep.,} vol.~4, 4139,
doi: 10.1038/srep04139.
\hfill\break

\noindent 
$\phantom{\,}$
\hskip-4.50truemm
Cai W, Chettiar U K, Kildishev A V, Shalaev V M (2007), 
``Optical cloaking with metamaterials,''
{\it Nature Photon.,} vol~1, pp.~224--227,
doi: 10.1038/nphoton.2007.28.
\hfill\break

\noindent 
$\phantom{\,}$
\hskip-4.50truemm
Chen H, Chen M (2011a), 
``Flipping photons backward: reversed Cherenkov radiation,''
{\it Materials Today,} vol.~14, pp.~34--41,
doi: 10.1016/S1369-7021(11)70020-7.
\hfill\break

\noindent 
$\phantom{\,}$
\hskip-4.50truemm
Chen J, Wang Y, Jia B, Geng T, Li X,  Feng L, Qian W,  Liang B, Zhang X, Zhuang S (2011b), 
``Observation of the inverse Doppler effect in negative-index materials at optical frequencies,''
{\it Nature Photon.,} vol.~5, pp.~239--242,
doi: 10.1038/nphoton.2011.130.
\hfill\break

$\phantom{\,}$
\hskip-4.50truemm
Cochran J F, Heinrich B, Dewar G (1977), 
``Ferromagnetic antiresonance transmission through Supermalloy at 24 GHz,''
{\it Can. J. Phys.,} vol.~55, pp.~787--805,
doi: 10.1139/p77-110.
\hfill\break

$\phantom{\,}$
\hskip-4.50truemm
Dewar G (2002), 
``The applicability of ferrimagnetic hosts to nanostructured
negative index of refraction (left-handed) materials,'' 
in Complex mediums III: beyond linear isotropic dielectrics, vol.~4806,
A.~Lakhtakia, G.~Dewar, and M.~McCall, Eds. 
Bellingham, Washington: SPIE,  pp.~156-166,
doi: 10.1117/12.472980.
\hfill\break

$\phantom{\,}$
\hskip-4.50truemm
Dewar G (2005a), 
``A thin wire array and magnetic host structure with $n<0$,''
{\it J. Appl. Phys.,} vol.~97, 10Q101,
doi: 10.1063/1.1846032.
\hfill\break

$\phantom{\,}$
\hskip-4.50truemm
Dewar G (2005b), 
``Minimization of losses in a structure having a negative index of refraction,''
{\it New J. Phys.,} vol.~7, 161,
doi: 10.1088/1367-2630/7/1/161.
\hfill\break

$\phantom{\,}$
\hskip-4.50truemm
Duan Z, Tang X,  Wang Z, Zhang Y, Chen X, Chen M,  Gong Y (2017), 
``Observation of the reversed Cherenkov radiation,''
{\it Nature Comm.,}
vol.~14, 14901,
doi: 10.1038/ncomms14901.
\hfill\break

$\phantom{\,}$
\hskip-4.50truemm
Gu Y, Wu R, Yang Y, Poo Y, Chen P, Lin Z (2013), 
``Self-biased magnetic left-handed material,''
{\it Appl. Phys. Lett.,} vol.~102, 231914,
doi: 10.1063/1.4811250.
\hfill\break

$\phantom{\,}$
\hskip-4.50truemm
Guenneau S, Puvirajesinghe T M (2013), 
``Fick's second law transformed: one path to cloaking in mass diffusion,''
{\it J. R. Soc. Interface,} vol.~10, 20130106,
doi: 10.1098/rsif.2013.0106.
\hfill\break

$\phantom{\,}$
\hskip-4.50truemm
He P, Gao J, Chen Y, Parimi P V, Vittoria C, Harris V G (2009), 
``Q-band tunable negative refractive index matamaterial using Sc-doped BaM hexaferite,''
{\it J. Phys. D: Appl. Phys.,} vol.~42, 155005,
doi: 10.1088/0022-3727/42/15/155005.
\hfill\break

$\phantom{\,}$
\hskip-4.50truemm
Leonhardt U,  Philbin T G (2007), 
``Quantum levitation by left-handed metamaterials,''
{\it New J. Phys.,} vol.~9, 254,
doi: 10.1088/1367-2630/9/8/254.
\hfill\break

$\phantom{\,}$
\hskip-4.50truemm
Moorish A H (1965), {\it The Physical Principles of Magnetism.}
New York: John Wiley \& Sons, pp.~607-616,
doi: 10.1002/9780470546581.
\hfill\break

$\phantom{\,}$
\hskip-4.50truemm
Pendry J B, Holden A J,  Stewart W J,  Youngs I (1996), 
``Extremely low frequency plasmons in metallic mesostructures,''
{\it Phys. Rev. Lett.,} vol.~76, pp.~4773--4776,
doi: 10.1103/PhysRevLett.76.4773.
\hfill\break

$\phantom{\,}$
\hskip-4.50truemm
Pendry J B (2000), 
``Negative refraction makes a perfect lens,''
{\it Phys. Rev. Lett.,} vol.~85, pp.~3966--3969,
doi: 10.1103/PhysRevLett.85.3966.
\hfill\break

$\phantom{\,}$
\hskip-4.50truemm
Pendry J B, D. Schurig D, Smith D R(2006), 
``Controlling Electromagnetic Fields,''
{\it Science,} vol.~312, pp.~1780--1782,
doi: 10.1126/science.1125907.
\hfill\break

$\phantom{\,}$
\hskip-5.50truemm
Pimenov A, Loidl A, Gehrke K, Moshnyaga V, Samwer K (2007), 
``Negative refraction observed in a metallic ferromagnet in the gigahertz frequency range,''
{\it Phys. Rev. Lett.,} vol.~98, 197401,
doi: 10.1103/PhysRevLett.98.197401.
\hfill\break

$\phantom{\,}$
\hskip-4.50truemm
Pokrovsky A L, Efros A L (2002), 
``Electrodynamics of metallic photonic crystals and the problem of left-handed materials,''
{\it Phys. Rev. Lett.} vol.~89, 093901,
doi: 10.1103/PhysRevLett.89.093901.
\hfill\break

$\phantom{\,}$
\hskip-4.50truemm
Rachford F J, Armstead D N, Harris V G,  Vittoria C (2007), 
``Simulations of Ferrite-Dielectric-Wire Composite Negative Index Materials,''
{\it Phys. Rev. Lett.,} vol~99, 057202,
doi: 10.1103/PhysRevLett.99.057202.
\hfill\break

$\phantom{\,}$
\hskip-5.50truemm
Restrepo-Fl\'orez J M, Maldovan M (2016), 
``Mass separation by metamaterials,''
{\it Sci. Rep.,} vol.~6, 21971,
doi: 10.1038/srep21971.
\hfill\break

$\phantom{\,}$
\hskip-4.50truemm
Schurig D, Mock J J, Justice B J, Cummer S A, Pendry J B, A. Starr A F, and
Smith D R(2006), 
``Metamaterial electromagnetic cloak at microwave frequencies,''
{\it Science,} vol.~314, pp.~977--980,
doi: 10.1126/science.1133628.
\hfill\break

$\phantom{\,}$
\hskip-4.50truemm
Shalaev V M (2007), 
``Optical negative-index metamaterials,''
{\it Nature Photon.,} vol.~1, pp.~41--48,
doi: 10.1038/nphoton.2006.49.
\hfill\break

$\phantom{\,}$
\hskip-4.50truemm
Shelby R A, Smith D R,  Schultz S (2001), 
``Experimental verification of a negative index of refraction,''
{\it Science,} vol.~292, pp.~77--79,
doi: 10.1126/science.1058847.
\hfill\break

$\phantom{\,}$
\hskip-4.50truemm
Smith D R, Padilla W J, Vier D C, Nemat-Nasser S C,  Schultz  S (2000), 
``Composite medium with simultaneously negative permeability and permittivity,''
{\it Phys. Rev. Lett.,} vol.~84, pp.~4184--4187,
doi: 10.1103/PhysRevLett.84.4184.
\hfill\break

$\phantom{\,}$
\hskip-4.850truemm
Veselago V G (1968), 
``The electrodynamics of substances with simultaneously negative values of $\epsilon$ and $\mu$,''
{\it Sov. Phys. USPEKHI,} vol.~10, pp.~517--526,
doi: 10.1070/pu1968v010n04abeh003699.
\hfill\break

$\phantom{\,}$
\hskip-4.50truemm
Wee W H, Pendry J B (2009), 
``Shrinking optical devices,''
{\it New J. Phys.,} vol.~11, 073033,
doi: 10.1088/1367-2630/11/7/073033.
\hfill\break

$\phantom{\,}$
\hskip-4.50truemm
Wood B (2009), 
``Metamaterials and invisibility,''
{\it Comptes Rendus Physique,} vol.~10, pp~379--390,
doi: 10.1016/j.crhy.2009.01.002.
\hfill\break

$\phantom{\,}$
\hskip-4.50truemm
Yang T, Chen H, Luo X, Ma H (2008),   
``Superscatterer: Enhancement of scattering with complementary media,''
{\it Optics Express,} vol.~16, pp.~18545--18550,
doi: 10.1364/OE.16.018545.
\hfill\break

$\phantom{\,}$
\hskip-4.0truemm
Zhao H, Zhou J, Zhao Q, Bo L,  Kang L (2007), 
``Magnetotunable left-handed material consisting of yittrium iron garnet slab and metallic wires,''
{\it Appl. Phys. Lett.,} vol.~91, 131107,
doi: 10.1063/1.2790500.
\hfill\break

$\phantom{\,}$
\hskip-4.50truemm
Zhao H, Zhou J, Kang L, Zhao Q (2009), 
``Tunable two-dimensional left-handed material consisting of ferrite rods and metallic wires,''
{\it Opt. Express,} vol.~17, pp.~13373--13380,
doi: 10.1364/OE.17.013373.
\hfill\break





\end{document}